# Mathematical problems on generalized functions and the canonical Hamiltonian formalism

J.F.Colombeau
jf.colombeau@wanadoo.fr

**Abstract.** This text is addressed to mathematicians who are interested in generalized functions and unbounded operators on a Hilbert space. We expose in detail (in a "formal way"- as done by Heisenberg and Pauli - i.e. without mathematical definitions and then, of course, without mathematical rigour) the Heisenberg-Pauli calculations on the simplest model close to physics. The problem for mathematicians is to give a mathematical sense to these calculations, which is possible without any knowledge in physics, since they mimick exactly usual calculations on $C^\infty$ functions and on bounded operators, and can be considered at a purely mathematical level, ignoring physics in a first step. The mathematical tools to be used are nonlinear generalized functions, unbounded operators on a Hilbert space and computer calculations. This text is the improved written version of a talk at the congress on linear and nonlinear generalized functions "Gf 07" held in Bedlewo-Poznan, Poland, 2-8 September 2007.

**1-Introduction**. The Heisenberg-Pauli calculations (1929) [We,p20] are a set of 3 or 4 pages calculations (in a simple yet fully representative model) that are formally quite easy and mimick calculations on $C^\infty$ functions. They are explained in detail at the beginning of this text and in the appendices. The H-P calculations [We, p20,21, p293-336] are a basic formulation in Quantum Field Theory: "canonical Hamiltonian formalism", see [We, p292] for their relevance. The canonical Hamiltonian formalism is considered as mainly equivalent to the more recent "(Feynman) path integral formalism": see [We, p376,377] for the connections between the 2 formalisms that complement each other.

  The H-P calculations start with irregular distributions that are "formally" multiplied, exponentiated,… Since 1955 it has been taught to physicists that mathematicians have "proved" (i.e. with a rigorous mathematical proof!) that there will never exist a nonlinear theory of generalized functions, in any mathematical context possibly different from distribution theory [S2,p10]. Thus this made hopeless to give the H-P calculations a rigorous mathematical sense. The above "proof" is discussed in this text.

  The H-P calculations make sense mathematically in the nonlinear theory of generalized functions: [C-G-P]. But only a too coarse work has been done. The H-P calculations request a deeper study: they concern generalized functions whose values are unbounded operators on a Hilbert space, which makes them more intricate than scalar valued calculations [C1,C2,C4], as done in continuum mechanics [C3,C8], and general relativity [S-V].

  Numerous presumably easy problems of pure mathematics and other ones presumably difficult are stated. Further, a reader interested in computer calculations can attempt the calculation of the new numerical predictions suggested by the mathematical understanding of the H-P formalism. Absolutely no knowledge of physics is needed. It suffices to know the definitions of generalized functions in the "simplified" or "special" case [G-K-O-S, p10] and to be interested in unbounded operators on a Hilbert space. [C-G-P] should be reconsidered in view of problems concerning domains of operators and numerical calculations (from imaginary exponentials of unbounded self-adjoint operators).

These problems should be solved before a presentation to theoretical physicists (i.e. mathematicians in pure mathematics and other ones in computer calculations should clarify simple mathematical models before one could propose theoretical physicists to reproduce their work on physically significant theories not accessible to mathematicians). Even if a future String Theory replaces Quantum Field Theory a mathematical understanding of QFT would be needed [Wi]. So the proposed works and problems are presumably important for the future. They are suited for mathematicians unaware of physics.

**2-The Heisenberg-Pauli calculations: overall view.** One has a specific Hilbert space called "the Fock space" denoted by **IF** (**IF** is a Hilbertian direct sum of spaces of square integrable functions, so a very familiar object for mathematicians, see Appendix 1 and [C-G-P]), and a family of explicitly defined objects $\Phi_0(x, t)$ ($x \in IR^3$, $t \in IR$) imagined to be linear operators on the Hilbert space **IF** for each (x,t) (see Appendix 1 and [C-G-P]). This family $\{\Phi_0(x, t)\}_{(x \in IR^3, t \in IR)}$ - equivalently the map $\Phi_0$ - is called "the free field operator". After 1955 it was recognized that the free field operator $\Phi_0$ is a distribution in the variable x (and a usual function in the t variable) whose "values" are densely defined unbounded linear operators on **IF**: more precisely there is a dense vector space **D** in **IF** such that, for any test function $\varphi$, $\Phi_0(\varphi, t)$ maps **D** into **D**. The H-P calculations, or "canonical Hamiltonian formalism", are described as follows on the simplest model close to physics. Let $\tau$, m, g $\in$ IR. One sets:

$$\mathbf{H}_0(\tau) := \int_{\xi \in IR^3} \{½ \cdot (\frac{\partial}{\partial t} \Phi_0(\xi, \tau))^2 + ½ \cdot \sum_{1 \leq \mu \leq 3} (\frac{\partial}{\partial x_\mu} \Phi_0(\xi, \tau))^2 + ½ \cdot m^2 \cdot (\Phi_0(\xi, \tau))^2 + g/(N+1) \cdot (\Phi_0(\xi, \tau))^{N+1}\} d\xi$$

where the powers mean composition of operators. For a mathematician the main point is that this formula involves products of irregular distributions (these products do not make sense within distribution theory): $\Phi_0(x,t)$ may be considered intuitively as a distribution as irregular as $\delta(x)$ the Dirac delta distribution (further, whose values are operators on a Hilbert space). This formula also involves an unjustified integration. Then one sets:

$$\Phi(x, t, \tau) := \exp(i(t-\tau)\mathbf{H}_0(\tau)) \circ \Phi_0(x, \tau) \circ \exp(-i(t-\tau)\mathbf{H}_0(\tau)).$$

Of course the exponentials are à fortiori not defined within the distributions: they are imagined as unitary operators on **IF** (because $\mathbf{H}_0(\tau)$ looks symmetric). There is also a problem of composition of operators because of $\Phi_0(x, \tau)$. Then calculations mimicking calculations on $C^\infty$ functions give that $\Phi(x,t,\tau)$ is solution of the Cauchy problem called "interacting field equation" (the "formal proof" is in section 3 below):

"**Theorem**"1. $\frac{\partial^2}{\partial t^2} \Phi(x, t, \tau) = \sum_{1 \leq \mu \leq 3} \frac{\partial^2}{\partial x_\mu^2} \Phi(x, t, \tau) - m^2 \Phi(x,t,\tau) - g \cdot (\Phi(x,t,\tau))^N$,

$\Phi(x, \tau, \tau) = \Phi_0(x, \tau)$,

$$\frac{\partial}{\partial t}\Phi(x,\tau,\tau) = \frac{\partial}{\partial t}\Phi_0(x,\tau).$$

It is a "wave" equation with nonlinear second member. Since the initial condition is a pair of irregular distributions the solution is not expected more regular than a distribution for which the nonlinear term does not make sense in distribution theory (with further a "big" problem due to the fact one is confronted with unbounded operators).

The exponential $\exp(it\mathbf{P}_0)$ of the energy operator $\mathbf{P}_0$ is immediately defined from an explicit formula that causes no mathematical problem: appendix 2 and [C-G-P]. It is a unitary operator on **IF** and it maps **D** into **D**. We set:

$$S_\tau(t) := \exp(i(t-\tau)\mathbf{P}_0) \circ \exp(-i(t-\tau)\mathbf{H}_0(\tau)).$$

Recall that the second exponential is not defined mathematically within the distributions. $S_\tau(t)$ is imagined as a unitary operator on **IF**. Then one has (from calculations mimicking calculations on functions: the "formal proofs" of th. 2 and 3 are in appendix 2):

"**Theorem" 2**.  $\Phi(x, t, \tau) = (S_\tau(t))^{-1} \circ \Phi_0(x, t) \circ S_\tau(t).$

This formula suggests that the numerical results of the theory are the limits when $\tau \to -\infty$, $t \to +\infty$ of the scalar products $|\langle F_1, S_\tau(t)F_2\rangle_{IF}|$, $F_1, F_2 \in \mathbf{D}$ (probability that an initial state $F_2$ become $F_1$ after interaction). One obtains (from formal calculations) that $S_\tau(t)$ is solution of the differential equation:

"**Theorem" 3.**  $\frac{d}{dt}S_\tau(t) = -i\frac{g}{N+1}\int_{\xi \in IR^3}(\Phi_0(\xi,t))^{N+1}d\xi \circ S_\tau(t)$

$S_\tau(\tau)$ =Identity operator.

This ordinary differential equation (in which $(\Phi_0(\xi,t))^{N+1}$ does not make sense in distribution theory) is the starting point of an attempt (called "perturbation theory") of calculation of $S_\tau(t)$ by developing it in powers of the "coupling constant" g, when g is small. Already before 1950 the results were formidable: for instance in the well known case of the anomalous magnetic moment of the electron the numerical value issued from the above calculations [with the coupled Maxwell-Dirac equations as interacting field equation – the H-P calculations are exactly similar to those in this paper–, and after other unjustified "manipulations" in a power series development (in powers of g)] is 1.00115965246 ± 20 in the last digits (i.e. $\pm 20.10^{-11}$; this imprecision is due to an imprecision on the physical constants like m, g above), while the experimental value (about 1950) was 1.00115965221 ± 4 in the last digits. This appears as the evidence that, although mathematically meaningless within distribution theory, the H-P calculations contain a "deep truth" in physics and, for those who believe mathematics and physics are connected, should make sense mathematically in some way.

**3-The Heisenberg-Pauli calculations: details.** In order to understand the H-P calculations we give the detailed calculations in proof of theorem 1: the passage from a formula to the following one is immediate. Concerning the free field it is convenient to set

$$\mathbf{\Pi}_0(x, t) := \frac{\partial}{\partial t} \Phi_0(x, t).$$

From the explicit formula of $\Phi_0$ one has the canonical commutation relations ($[A,B] := A \circ B - B \circ A$), see Appendix 1:

(0)
$$[\Phi_0(x, t), \Phi_0(x', t)] = 0,$$
$$[\mathbf{\Pi}_0(x, t), \mathbf{\Pi}_0(x', t)] = 0,$$
$$[\Phi_0(x, t), \mathbf{\Pi}_0(x', t)] = i\delta(x-x')\mathrm{Id},$$

with $\delta$ = the Dirac delta distribution. Then the calculations of the canonical Hamiltonian formalism run as follows. Set

(1) $\quad \mathbf{H}_0(\tau) := \int_{\xi \in \mathrm{IR}^3} \{½.(\mathbf{\Pi}_0(\xi,\tau))^2 + ½.\sum_{1 \leq \mu \leq 3} (\partial_\mu \Phi_0(\xi,\tau))^2 + ½.m^2.(\Phi_0(\xi,\tau))^2$

$+ g/(N+1).(\Phi_0(\xi,\tau))^{N+1} \} d\xi.$

Then set:

(2) $\quad \Phi(x, t) := \exp(i(t-\tau)\mathbf{H}_0(\tau)) \circ \Phi_0(x,\tau) \circ \exp(-i(t-\tau)\mathbf{H}_0(\tau))$

$\quad \mathbf{\Pi}(x, t) := \exp(i(t-\tau)\mathbf{H}_0(\tau)) \circ \mathbf{\Pi}_0(x,\tau) \circ \exp(-i(t-\tau)\mathbf{H}_0(\tau)).$

From (0) and (2) (simplifications of exponentials):

(3)
$$[\Phi(x, t), \Phi(x', t)] = 0,$$
$$[\mathbf{\Pi}(x, t), \mathbf{\Pi}(x', t)] = 0,$$
$$[\Phi(x, t), \mathbf{\Pi}(x', t)] = i\delta(x-x')\mathrm{Id}.$$

Formal differentiation of the exponentials in (2) gives:

(4)
$$\frac{\partial}{\partial t} \Phi(x, t) = i.[\mathbf{H}_0(\tau), \Phi(x, t)]$$
$$\frac{\partial}{\partial t} \mathbf{\Pi}(x, t) = i.[\mathbf{H}_0(\tau), \mathbf{\Pi}(x, t)].$$

Set:

(5) $\quad \mathsf{H}(\xi,t) := ½.(\mathbf{\Pi}(\xi,t))^2 + ½.\sum_{1 \leq \mu \leq 3} (\partial_\mu \Phi(\xi,t))^2 + ½.m^2.(\Phi(\xi,t))^2$

$+ g/(N+1).(\Phi(\xi,t))^{N+1}.$

(1),(2) and (5) imply (simplifications of exponentials):

(6) $\quad \int_{\xi \in \mathrm{IR}^3} \mathsf{H}(\xi,t) d\xi = \mathbf{H}_0(\tau).$

(4) and (6) imply :

(7)
$$\frac{\partial}{\partial t} \Phi(x, t) = i \int_{\xi \in \mathbb{R}^3} [\mathsf{H}(\xi,t), \Phi(x, t)] \, d\xi,$$

$$\frac{\partial}{\partial t} \Pi(x, t) = i \int_{\xi \in \mathbb{R}^3} [\mathsf{H}(\xi,t), \Pi(x, t)] \, d\xi.$$

From (5) and (3), (7) implies (use of the commutation relations (3) inside (7), from the explicit formula (5) of H):

(8)
$$\frac{\partial}{\partial t} \Phi(x, t) = \int_{\xi \in \mathbb{R}^3} \delta(\xi - x) \cdot \Pi(\xi, t) \, d\xi,$$

$$\frac{\partial}{\partial t} \Pi(x, t) = \int_{\xi \in \mathbb{R}^3} \{ -\sum_{1 \leq \mu \leq 3} \partial_\mu \Phi(\xi,t) \cdot \partial_\mu \delta(\xi - x) - m^2 \cdot \Phi(\xi,t) \cdot \delta(\xi - x)$$
$$- g \cdot (\Phi(\xi,t))^N \cdot \delta(\xi - x) \} \, d\xi.$$

(for the last equation use : $[\Phi^{N+1}, \Pi] = \Phi^N [\Phi, \Pi] + \Phi^{N-1}[\Phi, \Pi]\Phi + \ldots + [\Phi, \Pi]\Phi^N$, and from (3) $[\Phi, \Pi]$ is a multiple of the identity). Finally (8) gives the interacting field equation:

$$\frac{\partial}{\partial t} \Phi(x, t) = \Pi(x, t)$$

$$\frac{\partial}{\partial t} \Pi(x, t) = \sum_{1 \leq \mu \leq 3} \frac{\partial^2}{\partial x_\mu^2} \Phi(x, t) - m^2 \Phi(x, t) - g(\Phi(x,t))^N.$$

The detailed calculations for "theorems" 2 and 3 are of the same nature and not more difficult, see Appendix 2. Finally the requested level to "understand formally" the H-P calculations is the one of any undergraduate student in mathematics or physics: it suffices to imagine one computes on usual $\mathbf{C}^\infty$ functions of (x,t) whose pointvalues would be bounded operators on the Hilbert space (both assumptions are completely false: this difficulty has been left aside for mathematicians by Heisenberg–Pauli). Now a catastrophe!:

**4-"Multiplication of distributions is impossible in the general case, even in any theory, possibly different from distribution theory, where differentiation is always possible and that contains a delta function".** This strong claim in [S2,p10] presumably implies indeed that **the Heisenberg-Pauli calculations would never make sense mathematically**, since it is clear they deeply manipulate irregular distributions (the free field operator is indeed a rather irregular distribution, intuitively at least as irregular as the Dirac delta distribution) as if they were usual $\mathbf{C}^\infty$ functions. This strong claim is presented in [S2,p10] as an obvious consequence of a precise theorem [S1] that we state now:

**Theorem:** Let A be a algebra such that:
1) the algebra $C^0(\mathbb{R})$ (of all continuous functions on $\mathbb{R}$) is a subalgebra of **A**
2) the function $x \to 1$ is the unit element in A.
3) there exists a linear map $D: \mathbf{A} \to \mathbf{A}$, ("differentiation") such that

i) D reduces to the usual differentiation on $C^1$ functions

ii) $D(uv) = Du.v + u.Dv \quad \forall u,v \in \mathbf{A}$.

Then $D \circ D(x \to |x|) = 0$.

Since $D \circ D(x \to |x|)$ has to be $2\delta$, the algebra $\mathbf{A}$ is not acceptable as an algebra of generalized functions. The somewhat technical proof becomes extremely simple if 1) is replaced by: "the algebra spanned by the Heaviside function H is a subalgebra of $\mathbf{A}$". Then:

in the algebra $\mathbf{A}$ one has: $\quad H^2 = H$ and $H^3 = H$;

by differentiation (we note DH = H'): $\quad 2HH' = H'$ and $3H^2H' = H'$,

since $H^2 = H$ in the algebra $\mathbf{A}$: $\quad H^2H' = HH'$, thus $3HH' = H'$,

thus we have at the same time that $HH' = 1/2.H'$ and $HH' = 1/3.H'$, which implies $H' = 0$. ∎

**5-Let us try to understand what is behind this calculation.** Another quasi-identical calculation might be clearer. Let us consider the integral
$$I = \int (H^2(x) - H(x)).H'(x)dx.$$
H may be considered as an idealization of a $C^1$ function with a jump from the value 0 to the value 1 in a tiny interval around $x = 0$. Thus classical calculations are justified: $I = [H^3/3 - H^2/2]_{-\infty}^{+\infty} = 1/3 - 1/2 = -1/6$. This implies that $H^2 \neq H$ (since $I \neq 0$): $H^2$ and $H$ differ at $x = 0$, precisely where H' takes an "infinite value". This looks strange since in classical mathematics a function such as $H^2 - H$ is considered as null (from the viewpoint of integration theory). The classical formula $H^2 = H$ has to be considered as erroneous in a context suitable to compute I and one is inclined to state
$$H^2 - H = i,$$
with i some function considered as "infinitesimal" since it is null in classical mathematics, although it cannot be considered as null here. Then:
$$H^2 H' - H H' = i H',$$
and
$$I = \int i(x).H'(x)dx = -1/6:$$
iH' is not infinitesimal although i is! The replacement of : $H^2 = H$ and $H^3 = H$ by : $H^2 = H + i_1$ and $H^3 = H + i_2$, with $i_1, i_2$ "infinitesimal functions" suppresses the contradiction in the calculation at end of section 4, and the results of classical mathematics are recovered by replacing "infinitesimal function" by "zero function".

To summarize: in order to compute I one has been forced to introduce "infinitesimal functions" which are nonzero and which are considered as zero in classical mathematics at the price that then one cannot compute the integral I. In presence of these infinitesimal functions the proof of the impossibility theorem above fails.

26 years ago I introduced an algebra $\mathbf{G}(\Omega)$, $\Omega$ an open set in $\mathbb{R}^N$, of nonlinear generalized functions on $\Omega$ containing canonically the distributions ("full algebra" in [G-K-O-S]). The algebra $\mathbf{C}^\infty(\Omega)$ of all $\mathbf{C}^\infty$ functions on $\Omega$ is a subalgebra of $\mathbf{G}(\Omega)$. But from the theorem in section 4 the algebra $\mathbf{C}^0(\Omega)$ is not a subalgebra of $\mathbf{G}(\Omega)$. If f, g are 2 continuous functions on $\Omega$ their new product $f \bullet g$ in $\mathbf{G}(\Omega)$ differs from their classical product f.g in $\mathbf{C}^0(\Omega)$ by an "infinitesimal function" i (a concept rigorously defined in the $\mathbf{G}$-context):

$$f \bullet g = f.g + i :$$

$C^0(\Omega)$ is not a subalgebra of $G(\Omega)$, it is only a "**subalgebra modulo infinitesimals**". One has:

$$(f+i_1) \bullet (g+i_2) = f.g + i_3$$

where $i_1$, $i_2$ and $i_3$ are infinitesimal functions. If one drops all infinitesimals then one obtains the equality of the two products (usual terminology: two generalized functions that differ by an infinitesimal function are said to be "associated"; in the **G**-theory the word "associated" has been used instead of the more intuitive word "infinitesimal" that did not come to my mind when I found the **G**-theory). As exemplified in the above formula one observes that, in all calculations making sense within the distributions, dropping the infinitesimals gives nothing other than the classical calculations (the infinitesimals can -equally well- be dropped whenever they appear, or from time to time inside the calculations, or only at the end). In this way **the G-theory is perfectly coherent with classical mathematics and distribution theory.** But in **G-**calculations that do not make sense within the distributions, dropping the infinitesimals usually leads to nonsense as ascertained in calculations above.

**Problem.** Try to explain why the communities of mathematicians and mathematical physicists have widely accepted the Schwartz claim in title of section 4 without any criticism, since 1955, and why many of them (after 50 years!) still believe in this claim .

**Comments.** This might be connected with a rejection of "infinitesimals" from mathematics about 150 years ago. Another explanation might be that the multiplication of distributions is somewhat hidden. Here is the guideline that was used [C5, C7]. The reader is assumed – in this specific comments - to be aware of Schwartz notations: **D** denotes the space of all $C^\infty$ functions on $\Omega$ (any non void open set in $IR^n$) with compact support, **E'** denotes the space of all distributions on $\Omega$ with compact support. Let $C^\infty(D)$, respectively $C^\infty(E')$, denote the algebras of all $C^\infty$ functions on **D** and **E'** respectively (complex or real valued) [C9]. $\delta_x$ denotes as usual the Dirac delta distribution centered at the point x. One can use as well the holomorphic functions on **D** and **E'** instead of the $C^\infty$ functions [C10].

<u>First remark</u>: *$C^\infty(E')$ is contained in $C^\infty(D)$,* through the restriction map (which is injective since it is known from Schwartz distribution theory that **D** is contained and dense in **E'**).

<u>Second remark</u>: If $I = \{\Phi \in C^\infty(E')$ such that $\Phi(\delta_x)=0$ for all $x \in \Omega \}$, **I** is an ideal of $C^\infty(E')$ and *the quotient algebra $C^\infty(E')/ I$ is isomorphic as an algebra to the classical algebra of all $C^\infty$ functions on $\Omega$.*

<u>Thus the emergence of an idea serving as guideline</u>: *Try to extend the ideal I of $C^\infty(E')$ into an ideal J of $C^\infty(D)$, hoping that the elements of the quotient algebra $C^\infty(D)/ J$ would retain, from the particular case of $C^\infty(E')/ I$, properties of the $C^\infty$ functions on $\Omega$.* Modulo minor technical details this guideline worked and provided an algebra $G(\Omega)$ of nonlinear generalized functions on $\Omega$ [C5,C7]. Then the nonlinear theory of generalized functions was published thanks to support of L. Schwartz [C7] and L. Nachbin [C5,C6].

From 35 years ago to 26 years ago, before I found $G(\Omega)$, I had kept attempting to give a mathematical sense to the H-P calculations using $C^\infty$ or holomorphic functions on **D** [C9, C10] as an extension of the distributions although they do not enjoy enough properties of functions on $\Omega$ to be qualified as "generalized functions". The nonlinear generalized functions appeared as an attempt to improve the properties of the $C^\infty$ or holomorphic functions on **D** in view of the H-P calculations. Therefore I was not stopped by the Schwartz impossibility result. Further I was not worried by "infinitesimal quantities" since I was familiar with "infinite quantities" (their inverses) such as the "zero-point energy", those occurring in the Feynman graphs, or in functions of "infinite $L^2$ norm to be divided by this norm so as to produce a physical state". Then it appeared clear that classical continuum mechanics was far simpler than QFT for applications [C2,C3,C4,C8].

**6-The calculations of the canonical Hamiltonian formalism make sense in the G-setting,** see [C-G-P]. But they are made much more difficult than scalar calculations as in [C3, C8, G-K-O-S, S-V] from the fact they deal with generalized functions whose values are unbounded operators on a Hilbert space. Therefore there remain serious needs of improvements (by mathematicians) in unbounded operators (and also in computer calculations). It is the purpose of the sequel of this text to expose these problems. Of course one should keep in mind:

**Problem 1: modify the presentation in [C-G-P]** (considered as a basis for improvement following exactly the formal H-P calculations, in which –as well as in this text–there is some confusion between nonlinear generalized functions and their representatives to make these papers accessible to a reader not acquainted with nonlinear generalized functions) **so as to have a nicer presentation more suitable for the solution of the problems or for physical interpretation.**

The calculations presented in [C-G-P] consist in interpreting the free field operator $\Phi_0$ in the simplest possible way (i.e. as a distribution in the x-variable imbedded in the **G**-context by convolution) with representative $\Phi_0(\varphi,\varepsilon,x,t)$ (with the auxiliary function $\varphi$ such that $F\varphi$, the Fourier transform of $\varphi$, is $C^\infty$ with compact support in $\mathrm{IR}^3$ and identical to 1 on a 0-neighborhood in $\mathrm{IR}^3$). Each $\Phi_0(\varphi,\varepsilon,x,t)$ maps **D** into **D**. We set:

$$\mathbf{H}_0(\varphi,\varepsilon,\tau) = \int_{\xi \in \mathrm{IR}^3} \{½.(\partial_t \Phi_0(\varphi,\varepsilon,\xi,\tau))^2 + ½. \sum_{1\leq\mu\leq 3} (\partial_\mu \Phi_0(\varphi,\varepsilon,\xi,\tau))^2 +$$
$$½.m^2.(\Phi_0(\varphi,\varepsilon,\xi,\tau))^2 + g/(N+1).(\Phi_0(\varphi,\varepsilon,\xi,\tau))^{N+1}\} \cdot \chi(\varepsilon\xi)\,d\xi,$$

with $\chi$ a $C^\infty$ real valued function on $\mathrm{IR}^3$ with compact support and identical to 1 in a 0-neighborhood, so as to permit the integration. $\mathbf{H}_0(\varphi,\varepsilon,\tau)$ maps **D** into **D** and is symmetric. One can prove [C5 p311-313] it admits a self–adjoint extension denoted by $\mathbf{H}_0^\wedge$ on a domain $\mathbf{D}_0^\wedge$ containing **D** (do not forget $\mathbf{H}_0^\wedge$ and $\mathbf{D}_0^\wedge$ depend on $(\varphi,\varepsilon,\tau)$). Therefore, from the Hille-Yoshida theory, $\{\exp(it\,\mathbf{H}_0^\wedge)\}_{t \in \mathrm{IR}}$ is a strongly continuous group of unitary operators on **IF**. We set (representative of the interacting field operator):

(ifo)    $\Phi(\varphi,\varepsilon,\tau,x,t) := \exp(i(t-\tau)\mathbf{H}_0^\wedge) \circ \Phi_0(\varphi,\varepsilon,x,\tau) \circ \exp(-i(t-\tau)\mathbf{H}_0^\wedge)$

defined on $\mathbf{D}(t,\varphi,\varepsilon,\tau) := \exp(i(t-\tau)\mathbf{H}_0^\wedge)\mathbf{D}$, which is a dense subspace of **IF,** a priori depending on t (and also on $(\varphi,\varepsilon,\tau)$), but independent on x.

**7- The problem of domains.** From (ifo), $\Phi(\varphi,\varepsilon,\tau,x,t)$ maps $\mathbf{D}(t,\varphi,\varepsilon,\tau)$ into $\mathbf{D}(t,\varphi,\varepsilon,\tau)$, and so does its expected time-derivative:

$$i\,\mathbf{H}_0^\wedge \circ \exp(i(t-\tau)\mathbf{H}_0^\wedge) \circ \Phi_0(\varphi,\varepsilon,x,\tau) \circ \exp(-i(t-\tau)\mathbf{H}_0^\wedge) +$$
$$\exp(i(t-\tau)\mathbf{H}_0^\wedge) \circ \Phi_0(\varphi,\varepsilon,x,\tau) \circ \exp(-i(t-\tau)\mathbf{H}_0^\wedge) \circ (-i)\mathbf{H}_0^\wedge =$$
$$i\,\exp(i(t-\tau)\mathbf{H}_0^\wedge) \circ \mathbf{H}_0 \circ \Phi_0(\varphi,\varepsilon,x,\tau) \circ \exp(-i(t-\tau)\mathbf{H}_0^\wedge) +$$
$$\exp(i(t-\tau)\mathbf{H}_0^\wedge) \circ \Phi_0(\varphi,\varepsilon,x,\tau) \circ (-i)\,\mathbf{H}_0 \circ \exp(-i(t-\tau)\mathbf{H}_0^\wedge) =$$
$$i\,\exp(i(t-\tau)\mathbf{H}_0^\wedge) \circ [\,\mathbf{H}_0, \Phi_0(\varphi,\varepsilon,x,\tau)\,] \circ \exp(-i(t-\tau)\mathbf{H}_0^\wedge).$$

In order to define the t-derivative of $\Phi(\varphi,\varepsilon,\tau,x,t)$ it would be desirable to have a domain independent on t.

**Problem 2: is it possible to find a dense vector subspace of IF, independent on t (as well of x, of course), that could be a domain of definition of $\Phi(\varphi,\varepsilon,\tau,x,t)$ and permit a nice interpretation of the t-derivative? Then in what sense would $\Phi(\varphi,\varepsilon,\tau,x,t)$ satisfy properties similar to those listed in** [Jo, p52-55] **to be satisfied by an interacting field operator?**

The H-P calculations that we followed exactly in [C-G-P] give $\Phi$ as a generalized function in x and a classical function in t. Physicists check formally that $\Phi$ is Lorentz invariant. Therefore it is likewise that $\Phi$ should better be considered in problem 2 as a generalized function in (x,t). Could the method in [Ja chap7, p91-103, G-J] give the desired result? Note that this problem of domains appears already in ordinary quantum mechanics (with a finite number of degrees of freedom) in absence of generalized functions [K p1-4].

In order to solve this problem one might also attempt to modify the starting point, i.e. the formula of $\Phi_0(\varphi,\varepsilon,x,t)$ (which was chosen in [C-G-P] only as the "most natural one" on an argument of simplicity); for instance replace $\Phi_0(\varphi,\varepsilon,x,t)$ by its Yoshida regularisation for the value $\varepsilon$ of the parameter. Then the problem comes from the fact this might –but only "infinitesimally" – modify the commutation relations [§3 (0)].

**Problem 3: does this modified choice of free field operator in the G-context lead to the interacting field equation at least in the association sense? Try also possible other choices.**

Heisenberg and Pauli compute formally on unbounded linear operators as if they were bounded operators. Kernels of (densely defined unbounded) linear operators on the Hilbert space $L^2(IR^n)$ have been introduced in [B-C-D1, B-C-D2, De] in order to define a composition product for operators that cannot be composed in the usual sense, so as to avoid the above problem of domains. Kernels of (in general unbounded) operators on the Hilbert space $L^2(IR^n)$ are nonlinear generalized functions on $IR^n \times IR^n$ and one has the usual integral formula connecting an operator and its kernel: if O is a linear operator on $L^2(IR^n)$ its kernel $K \in \mathbf{G}(IR^n \times IR^n)$ is defined from the formula
$$(O(f))(x) = \int K(x,y) f(y) dy$$
and the composition of operators is then defined through the formula
$$(O_2 \circ O_1(f))(x) = \int K_2(x,\xi) K_1(\xi,y) f(y) d\xi dy$$

which makes sense in the **G**-context if $K_1, K_2 \in \mathbf{G}(IR^n \times IR^n)$ provided the integral is convergent. The Schwartz kernel theorem asserts that "any usual", even unbounded, linear operator on $L^2(IR^n)$ has a kernel which is a distribution on $IR^n \times IR^n$. Therefore the products of these kernels make sense in the **G**-context. That is why the **G**-context provides a definition of the composition product of unbounded operators in cases in which this is not possible classically. Of course this concept also opens research tracks independently of QFT.

**Problem 4: Use the kernels to justify the t-derivation above.** The kernels could be used at 2 levels: only to avoid the dependence of the domain of $\Phi(\varphi, \varepsilon, \tau, x, t)$ by a definition of composition that would not involve the domains, or from the beginning by trying to define the exponentials via the kernels by some arguments similar to those in the Hille-Yoshida theory (I failed to define the needed exponentials from kernels by adaptation of the usual series). One can try to transfer the situation to $L^2(IR^3)$ or develop the kernel theory in **IF**.

**8- Numerical results.** The final numerical results $|\langle F_1, S_\tau(t) F_2 \rangle_{IF}|$, $F_1, F_2 \in \mathbf{D}$, appear as "limits" (in some sense to be clarified) when $t \to +\infty, \tau \to -\infty$ of "bounded oscillating" generalized real numbers: for fixed t, $\tau$ the representatives $|\langle F_1, S_\tau(t) F_2 \rangle_{IF}|$ are bounded by $\|F_1\|_{IF} \cdot \|F_2\|_{IF} = 1$ (physical states have norm = 1) and presumably oscillate when $\varepsilon \to 0^+$, [C-G-P], and we have to associate to them ordinary real numbers. The concept of association used till now in the context of nonlinear generalized functions consists of letting $\varepsilon \to 0^+$ and finding a limit independent on the particular functions $\varphi$ in use. Here it seems doubtful that such a limit would exist and a more general definition of association is needed which should be intuitively of a "probabilistic nature", from an average on a "very large" number of "experiments" which appear mathematically as arbitrary choices of "very small" values of $\varepsilon > 0$ and arbitrary choices of convenient auxiliary functions $\varphi$.

Therefore a natural <u>probabilistic</u> interpretation of the association (of a classical real number to a generalized number) can be introduced by an averaging process (see [Ba, Be] for averaging results).

<u>Examples:</u> the generalized real number having as representative $R(\varphi, \varepsilon) := |\cos(g/\varepsilon)|$, $g \neq 0$ depending or not on $\varphi$, is oscillating endlessly as $\varepsilon \to 0^+$. One naturally associates to it an average value such as for instance the limit when $\eta \to 0^+$ of $\frac{1}{\eta} \int_0^\eta |\cos(g/\varepsilon)| d\varepsilon$ which can be proved to exist and $= 2/\pi = 0.636...$ . Idea of proof: it shows successive archs that "tend to" archs of the function $|\cos|$ as $\varepsilon \to 0^+$, compressed in the "horizontal" direction. The average value does not exist if $R(\varphi, \varepsilon) := |\cos(\log(1/\varepsilon))|$; then the predictive value is limited to an interval between the inf-limit and the sup-limit, in this case a very poor result: [0.44, 0.82] = $0.63 \pm 0.19$ (from numerical calculations). If $R(\varphi, \varepsilon) := |\cos(\log(1/\varepsilon^p))|$, then the uncertainty tends to 0 as $p \to +\infty$, which shows that the existence of an average value is not indispensable in a physical situation which can be modelled by $\varepsilon^p$ instead of $\varepsilon$ (usually OK).
An average on a random choice of a very large number of very small values of $\varepsilon$ gives also same results (this can easily be checked by numerical calculations), as well as slightly different integral formulas [C-G-P] whose aim is to capture the oscillations when $\varepsilon \to 0^+$ and take an average of them. It is proved in [Da] that, if f is any almost periodic function (uap function in the sense of [Be]) and if p >0 (p∈ IR), then $\frac{1}{\eta}\int_0^\eta f((1/\varepsilon)^p)d\varepsilon$ tends to a limit independent on p as $\eta \to 0^+$ (this is important because one could have chosen $\varepsilon^p$ as well as $\varepsilon$ in the formula defining the representative of the free field operator in [C-G-P]).

The second exponential in the formula $S_\tau(t) := \exp(i(t-\tau)\mathbf{P}_0) \circ \exp(-i(t-\tau)\mathbf{H}_0^\wedge)$ makes the problem look like the above |cos| examples, but much more technical. Unfortunately numerical calculations of $\exp(-i(t-\tau)\mathbf{H}_0^\wedge)$ appear difficult; from the Hille-Yoshida theory it is limit of $(I+ (i/n)(t-\tau)\mathbf{H}_0^\wedge)^{-n}$ as $n \to \infty$. One might also attempt a numerical solution from the ODE in th. 3: for given $F_2 \in \mathbf{D}$, $(S_\tau(\varphi,\varepsilon,t)F_2)$ is solution of an ODE valued in **IF**.

**Problem 5: prove the existence of such a probabilistic limit (or average value) independent on $\varphi$ and compute it numerically** (a calculation of inf-limit and sup-limit would already give a significative result provided these limits are close enough).

The numerical calculations might be significant even for a "nonrenormalizable theory" [We]; indeed a nonrenormalizable theory looks roughly like the "formal meaningless limit" when $\varepsilon \to 0^+$ of the expansion $|\cos(g/\varepsilon)| = |1 - (g/\varepsilon)^2/(2!) + (g/\varepsilon)^4/(4!) + ...|$, whose associated value (by averaging) is $2/\pi$.

If one develops $\langle F_1, S_\tau(t)F_2 \rangle_{IF}$ in powers of g (no convergence result has been proved so this is only a "formal" power series) one obtains the (analog of the) usual (nonrenormalized) perturbation series, but with coefficients generalized numbers, i.e. having representatives depending on $\varphi$ and $\varepsilon$.

**Problem 6: In case of a renormalizable theory ($N \leq 3$) try to prove that the averaged limit of $|\langle F_1, S_\tau(t)F_2 \rangle_{IF}|$ is approximated by the renormalized perturbation series for values of g close to 0.** Do it first on simplified mathematical examples.

The interpretation of renormalization has not been clarified: a possible idea would be (for fixed $\tau$,t to simplify) that the global result $\langle F_1, S_\tau(t)F_2 \rangle_{IF}$ would have an asymptotic expansion in powers of $\varepsilon$ if $\varepsilon \to 0^+$, and that each individual power of g would also have an asymptotic expansion starting with negative powers of $\varepsilon$, all of them $\geq -N_0$ ( some $N_0 \in IN$ related to the degree of divergence of the renormalizable theory). The coefficients of the negative powers of $\varepsilon$ in $\langle F_1, S_\tau(t)F_2 \rangle_{IF}$ would be of course all globally null because of the global boundedness of $\langle F_1, S_\tau(t)F_2 \rangle_{IF}$ as $\varepsilon \to 0^+$, and one might think that renormalization might consist in "killing" separately in the individual powers of g the participation of these globally null coefficients. For asymptotic expansions of nonlinear generalized functions see [C6, p78-93].

**General research track:** The **G**-context puts in evidence "abstract objects" (generalized functions or numbers) that are solution of equations. Then the genuine difficulty is shifted to the final task of ascertaining whether the solutions thus obtained are classical objects. This is done through the "association" process. The existence of an associated real or complex number in the above sense of some average as $\varepsilon \to 0^+$ (to be independent on $\varphi$) is considerably weaker than a limit in the usual sense as used up to now in the applications of nonlinear generalized functions (see [Be, Ba]: most types of almost periodic functions in the variable $1/\varepsilon$ do enjoy such an average, and such an average exists for lot of non almost periodic functions). <u>The use of this concept (of association through averaging) in the (numerous) works already done in the **G**-context will certainly bring lot of new results of association of classical objects to the "abstract" solutions found in the **G**-context</u>. Other averages such as $\frac{1}{\eta}(\int_0^\eta R(\varphi,\varepsilon)^2 d\varepsilon)^{1/2}$, if $R \geq 0$, might be considered.

**9-Final comments.** It is well known now, at least among the participants in this congress, that a convenient nonlinear theory of generalized functions does exist (recent surveys [C1,C2,C3,S-V]; see also [C4]). Therefore it becomes possible for mathematicians interested in unbounded operators on a Hilbert space to give a (more or less satisfactory) mathematical sense to the canonical Hamiltonian formalism, by trying to solve the problems listed above and trying to explore the numerous tracks stemming from possible modifications of [C-G-P] which can serve as a starting point. The difficulty might rather be that one faces too many possible research tracks. The needed prerequisites to start this research are quite reduced and practically contained in this paper and [C-G-P].

**Appendix 1.** We give preliminary definitions that are not in the main text.
**\*The Fock space.** The Fock space is the Hilbertian direct sum

$$\mathbf{IF} = \bigoplus_{n=0}^{+\infty} L_S^2((IR^3)^n),$$

where, for n>0, $L_S^2((IR^3)^n)$ is the Hilbert space of complex valued symmetric square integrable functions on $(IR^3)^n$ and for n=0 $L_S^2((IR^3)^n)$ stands for the field of complex numbers. That is: an element of **IF** is an infinite sequence $F = (f_n)_n$, n=0,…,∞, such that $|f_0|^2 + \sum_{n=1}^{+\infty}(\|f_n\|_n)^2 < +\infty$, where $\| \|_n$ is the norm in $L_S^2((IR^3)^n)$ and $f_n$ stands for the symmetric function $(k_1,...,k_n) \to f_n(k_1,...,k_n)$, $k_i \in IR^3$. Of course from the definition of a Hilbertian direct sum:

$$\|(f_n)\|_{IF} = (|f_0|^2 + \sum_{n=1}^{+\infty}(\|f_n\|_n)^2)^{1/2},$$

$$\langle (f_n),(g_n) \rangle_{IF} = f_0 \cdot \overline{g_0} + \sum_{n=1}^{\infty} \langle f_n, g_n \rangle_n$$

where $\langle , \rangle_n$ is the scalar product in $L_S^2((IR^3)^n)$. We shall use a dense subspace of **IF**:

$\mathbf{D} = \{(f_n)_n \in \mathbf{IF}$ such that $f_n = 0$ for n large enough$\}$.

**\*The creation and annihilation operators.** If $\psi \in L^2(IR^3)$ the creation operator $a^+(\psi)$ is given by the formula:

$$(f_n)_n \to (0, f_0 \psi,..., \sqrt{n}\, \text{Sym}(\psi \otimes f_{n-1}),...),$$

where Sym is the symmetrization operator. If $\psi \in L^2(IR^3)$ the annihilation operator $a^-(\psi)$ is given by the formula:

$$(f_n)_n \to (\int f_1(\lambda)\psi(\lambda)d\lambda, ..., \sqrt{n+1}\int f_{n+1}(.,\lambda)\psi(\lambda)d\lambda, ...).$$

The operators $a^+(\psi)$ and $a^-(\psi)$ are defined at least on the dense subspace **D** of **IF**, with values in **D**. They are not bounded operators on **IF** because of the coefficients $\sqrt{n}$ and $\sqrt{n+1}$.

**\*The free field operator.** The free field operator is defined by:

$$\Phi_0(x, t) = a^+(k \to 2^{-1/2}(2\pi)^{-3/2}(k^0)^{-1/2}e^{ik^0 t}e^{-ikx}) + a^-(k \to 2^{-1/2}(2\pi)^{-3/2}(k^0)^{-1/2}e^{-ik^0 t}e^{ikx}),$$

$k \in IR^3$, $k^0 = (k^2 + m^2)^{1/2}$. The functions of the variable $k$ inside $a^+$ and $a^-$ are not in $L^2(IR^3)$, so that the mathematical objects that make sense in distribution theory are:

$$\Phi_0(\varphi, t) := \int_{IR^3} \Phi_0(x, t)\,\varphi(x)dx,$$

with $\varphi$ a suitable test function. That is: $\Phi_0(x, t)$ is a distribution in the x-variable whose values $\Phi_0(\varphi, t)$ are densely defined linear unbounded operators on **IF** (they map **D** into **D**). We set: $\Pi_0(x, t) := \frac{\partial}{\partial t}\Phi_0(x, t)$, which is again a similar distribution. From the explicit formulas of $\Phi_0(x, t)$ and $\Pi_0(x, t)$ one obtains easily the commutation relations [§3 (0)] (one first checks the relations $[a^+(\psi), a^+(\psi')] = 0 = [a^-(\psi), a^-(\psi')]$ and $[a^-(\psi), a^+(\psi')] = \int \psi.\psi' dk$ .Id, then one uses the formula of $\Phi_0(x, t)$).

**Appendix 2.** <u>Proof of Th. 2</u>. The exponential $\exp(it\,\mathbf{P}_0)$ of the energy operator $\mathbf{P}_0$ is defined by:

$$(f_n)_n \to (f_0, k \to \exp(ik^0 t).f_1(k), ..., (k_1, ..., k_n) \to \exp(i(k_1^0 + ... + k_n^0)t).f_n(k_1, ..., k_n), ...).$$

It is a unitary operator on IF and it maps **D** into **D**. One checks easily that:

(transl) $\quad \Phi_0(x, t+\theta) = \exp(i\theta\,\mathbf{P}_0) \circ \Phi_0(x, t) \circ \exp(-i\theta\,\mathbf{P}_0)$, and same for $\Pi_0$.

Then
$$(S_\tau(t))^{-1} \circ \Phi_0(x, t) \circ S_\tau(t) =$$
$\exp(i(t-\tau)\mathbf{H}_0(\tau)) \circ \exp(-i.(t-\tau)\mathbf{P}_0) \circ \Phi_0(x,t) \circ \exp(i.(t-\tau)\mathbf{P}_0) \circ \exp(-i(t-\tau)\mathbf{H}_0(\tau)) =$
$\exp(i(t-\tau)\mathbf{H}_0(\tau)) \circ \Phi_0(x, t-(t-\tau)) \circ \exp(i(t-\tau)\mathbf{H}_0(\tau)) = \Phi(x, t, \tau)$,
from (transl) and the definition of $\Phi(x, t, \tau)$. We have obtained the formula in th.2. ∎

<u>Proof of Th. 3</u>. Derivation in t of
$$S_\tau(t) := \exp(i.(t-\tau)\mathbf{P}_0) \circ \exp(-i(t-\tau)\mathbf{H}_0(\tau))$$
gives:
$\frac{d}{dt}S_\tau(t) = i\,\mathbf{P}_0 \circ S_\tau(t) + S_\tau(t) \circ (-i\mathbf{H}_0(\tau)) =$
$i\,\mathbf{P}_0 \circ S_\tau(t) + \exp(i.(t-\tau)\mathbf{P}_0) \circ \exp(-i(t-\tau)\mathbf{H}_0(\tau)) \circ (-i\mathbf{H}_0(\tau)) =$

i $\mathbf{P}_0 \circ \mathbf{S}_\tau(t) + \exp(i.(t-\tau)\mathbf{P}_0) \circ (-i\mathbf{H}_0(\tau)) \circ \exp(-i(t-\tau)\mathbf{H}_0(\tau)) =$
$i\mathbf{P}_0 \circ \mathbf{S}_\tau(t) - i.\exp(i.(t-\tau)\mathbf{P}_0) \circ \mathbf{H}_0(\tau) \circ \exp(-i.(t-\tau)\mathbf{P}_0) \circ \exp(i.(t-\tau)\mathbf{P}_0) \circ \exp(-i(t-\tau)\mathbf{H}_0(\tau))$ $= i\mathbf{P}_0 \circ \mathbf{S}_\tau(t) - i \exp(i.(t-\tau)\mathbf{P}_0) \circ \mathbf{H}_0(\tau) \circ \exp(-i.(t-\tau)\mathbf{P}_0) \circ \mathbf{S}_\tau(t) =$
$- i \{ -\mathbf{P}_0 + \exp(i.(t-\tau)\mathbf{P}_0) \circ \mathbf{H}_0(\tau) \circ \exp(-i.(t-\tau)\mathbf{P}_0) \} \circ \mathbf{S}_\tau(t)$.

From (transl)  $\exp(i.(t-\tau)\mathbf{P}_0) \circ \mathbf{H}_0(\tau) \circ \exp(-i.(t-\tau)\mathbf{P}_0)) = \mathbf{H}_0(t)$ ;

thus one obtains

$$\frac{d}{dt} \mathbf{S}_\tau(t) = -i(-\mathbf{P}_0 + \mathbf{H}_0(t)) \circ \mathbf{S}_\tau(t).$$

Cumbersome calculations ( done in [C5, p22-24] ) on the first 3 types of terms in $\mathbf{H}_0(t)$ give a simplification of their sum with $\mathbf{P}_0$ and one obtains

$$\frac{d}{dt} \mathbf{S}_\tau(t) = -i \left( \int_{k \in IR^3} k^0 dk + \frac{g}{N+1} \int_{\xi \in IR^3} (\Phi_0(\xi,t))^{N+1} d\xi \right) \circ \mathbf{S}_\tau(t).$$

$\int_{k \in IR^3} k^0 dk$ is a first "infinity" encountered in the formal calculations. It is called the "zero point energy". It is suppressed because it does not influence the transition probabilities (from the absolute value in $|\langle F_1, S_{-\infty}(+\infty)F_2 \rangle_{IF}|$), and one obtains the formula in th 3. ∎